\documentclass{appolb}
\usepackage[utf8]{inputenc}
\usepackage{graphicx}
\usepackage{amsmath}
\usepackage{hyperref}

\newcommand{\tref}[1]{Tab.~\ref{#1}}
\newcommand{\fref}[1]{Fig.~\ref{#1}}

\begin{document}

\title{The quark propagators of QCD and QCD-like theories
\thanks{Presented at Excited QCD 2017, 7-13 May 2017, Sintra, Portugal}
}
\author{Romain~Contant, Markus~Q.~Huber
\address{Institute of Physics, University of Graz, NAWI Graz, Universit\"atsplatz 5, 8010 Graz, Austria}
}

\maketitle

\begin{abstract}
We investigate the phase structures of theories which differ from QCD only in the gauge group and can be simulated on the lattice at non-vanishing chemical potential $\mu$.
These theories can thus serve as testing ground for functional methods at non-vanishing density.
We determine the chiral and confinement/deconfinement transitions at $\mu=0$ for the three gauge groups $SU(3)$, $SU(2)$ and $G_2$ for two quark flavors and extend the study of the chiral transition to non-zero $\mu$.
We locate the critical point where the chiral crossover becomes a real phase transition.
Within the employed truncation, we find that all three theories behave qualitatively very similarly.
\end{abstract}

  
\section{Introduction}
\label{sec:introduction}

A worldwide effort is devoted to the study of the phase diagram of QCD both theoretically and experimentally. However, due to the complex action problem, Monte-Carlo simulations are currently not possible at large chemical potential~$\mu$ \cite{deForcrand:2010ys}. 
Functional methods like Dyson Schwinger Equations (DSEs) \cite{Alkofer:2000wg,Alkofer:2008nt} or functional renormalization group equations \cite{Berges:2000ew,Gies:2006wv} provide an alternative framework to explore the non-perturbative regime of quantum field theories.
DSEs are the equations of motions of the correlation functions of a quantum field theory. These non-perturbative equations consist of an infinite system of coupled (non-)linear equations. Thus, truncations are mandatory to solve them numerically. The pressing question is, of course, how well a truncation describes the underlying physics.
The most advanced truncations are actually able to describe the correlations functions of QCD and Yang-Mills theory quite well \cite{Cyrol:2016tym,Cyrol:2017ewj}.
In this work, we will study the effect of the medium on the matter sector of QCD and QCD-like theories within the DSE framework. For this purpose, we will extract the chiral and (at vanishing chemical potential) the confinement/deconfinement transitions from the corresponding quark propagators.
The different gauge groups studied are $SU(3)$ and $SU(2)$ and the exceptional group $G_2$. The last two do not suffer from the sign problem \cite{Kogut:2000ek,Holland:2003jy} and can be simulated at finite $\mu$ on the lattice, e.g., \cite{Maas:2012wr,Cotter:2012mb,Boz:2013rca,Braguta:2016cpw}.
They are in many respects similar to QCD, e.g., for the quenched theories the chiral and deconfinement transitions occur at the same critical temperatures. Moreover, the correlations functions as obtained with lattice methods are qualitatively very similar \cite{Maas:2007af,Sternbeck:2007ug,Cucchieri:2007zm,Fischer:2010fx,Maas:2011ez}. Thus, understanding the effects of truncations of functional equations in these QCD-like theories, we hope to learn also something about the equations in QCD.

\section{Setup}
\label{sec:setup}

\begin{figure}[tb]
\begin{center}
\includegraphics[width=6cm]{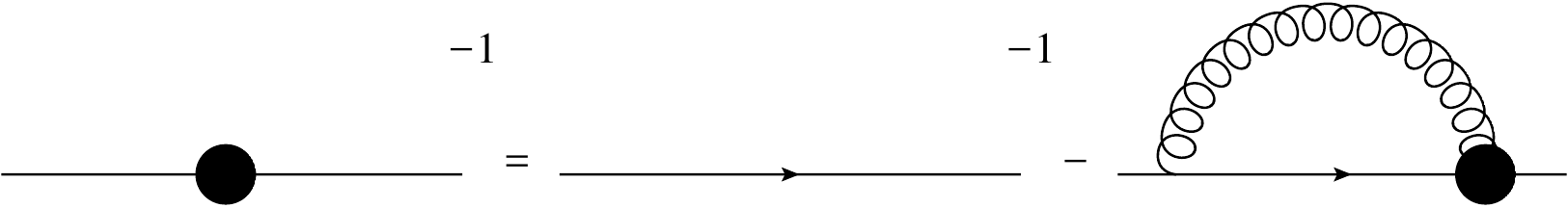}
\hfill
\includegraphics[width=6cm]{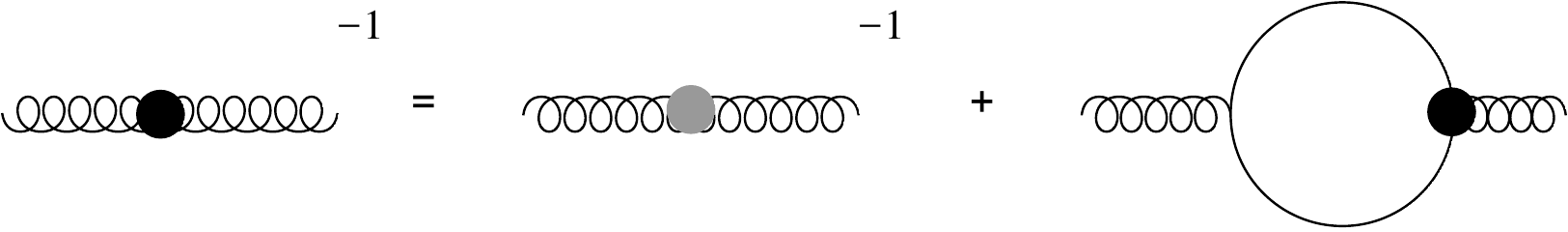}
\caption{The system of solved DSEs. Quantities with a black blob are fully dressed, as are internal propagators. Continuous/wiggly lines denote quarks/gluons. The gray blob denotes the approximated quenched part of the gluon propagator.}
\label{Fig:DSEs}
\end{center}
\end{figure}

At finite temperature and density one can write the quark propagator $S(\vec{p}, \omega_{n})$ with the following four dressing functions :
\begin{align}
S^{-1}(\vec{p}, \omega_{n}) = i \vec{p} \vec{\gamma} A(\vec{p}, \omega_{n}) + i \omega_{n} \gamma_{4} C(\vec{p}, \omega_{n})+ B(\vec{p}, \omega_{n}) +  i \omega_{n} \gamma_{4} \vec{p} \vec{\gamma} D(\vec{p}, \omega_{n}),
\end{align}
where $\omega_n = 2 \pi T (n + 1)$.
$D(\vec{p}, \omega_{n})$ vanishes in certain asymptotic cases and in Ref.~\cite{Contant:2017gtz} we showed explicitly that the contribution of $D(\vec{p}, \omega_{n})$ stays small for $\mu=0$ and $T\neq 0$. Hence it will be neglected in this work.
The dressing functions are calculated from the gap equation depicted in \fref{Fig:DSEs}. 

The chiral and confinement/deconfinement transitions are extracted from the chiral condensate $\left<\overline{\psi} \psi \right>_\varphi$ evaluated with a $U(1)$-valued boundary condition $\omega_n(\varphi) = 2 \pi T (n + \frac{\varphi}{2 \pi})$, $\varphi \in [0 , 2 \pi]$. At $\varphi = \pi$, the usual chiral condensate $\left<  \overline{\psi} \psi \right>$ is recovered which can be used to identify the chiral transition.
The Fourier transform of the $\varphi$-dependent chiral condensate w.r.t. $\varphi$ is called the dual chiral condensate $\Sigma$. It transforms under center transformations in the same way as the Polyakov loop and can thus be used as an order parameter for the quark confinement/deconfinement transition \cite{Bilgici:2008qy,Synatschke:2007bz,Fischer:2009wc}. The condensates are calculated as
\begin{align}
\Delta_{l,h} = -\left<  \overline{\psi} \psi \right>_{l} + \frac{m_l}{m_h}\left<  \overline{\psi} \psi \right>_{h},
\qquad \Sigma=\int_0^{2\pi}\frac{d\varphi}{2\pi}e^{-i\,\varphi}\left<  \overline{\psi} \psi \right>_\varphi d\varphi ,
\end{align}
where the quadratically divergent chiral condensate is regularized by subtracting the condensate with a heavier bare mass $m_h$ from the condensate of the light bare mass $m_l$.
The crossover temperatures are determined via the extrema of the derivatives of the condensates: 
\begin{align}
 \chi_{ch} = \frac{\partial \Delta_{l,h}}{\partial T},\qquad \chi_{dec} = \frac{\partial \Sigma}{\partial T}.
\end{align}

\label{sec:transition}

As input for the quark propagator DSE we need the gluon propagator dressing functions and the quark-gluon vertex. For the transverse and longitudinal gluon dressing functions we use \cite{Fischer:2010fx}
\begin{align}
 Z_{ T/L }(p^2)&=\frac { x }{ (x+1)^2 } \Bigg( \left( \frac { c/\Lambda^2 }{ x+a_{ T/L } }  \right) ^{ b_{ T/L } }+
 x\left( \frac { \alpha (\mu )\beta _{ 0 } }{ 4\pi  } \textrm{ln}(x+1)) \right) ^{ \gamma  } \Bigg),
\end{align}
where $x=p^2/\Lambda^2$, $\gamma$ is the anomalous dimension of the gluon, $\beta_0$ is the lowest coefficient of the $\beta$ function, $\alpha(\mu)$ is the coupling and the parameters $c = 11.5\,\text{GeV}^2$ and $\Lambda = 1.4 \,\text{GeV}$ are fixed. The temperature dependence enters via $a_{ T/L }$ and $b_{ T/L }$, which is determined by fits to quenched lattice data \cite{Fischer:2010fx,Maas:2011ez}.
The effects of the quarks in the gluon dressing will be added through the explicit calculation of the quark loop as introduced in \cite{Fischer:2012vc}. Fig.~\ref{Fig:DSEs} shows the complete system of DSEs we solve using the framework of \textit{CrasyDSE} \cite{Huber:2011xc}.

For the quark-gluon vertex we will use a model that effectively captures the infrared contribution in a dressing of the tree-level tensor $\gamma_{\mu}$ \cite{Fischer:2009gk}: 
\begin{align}
\Gamma_{\nu}(q;p,l) &= \gamma_{\mu} \Gamma_{mod}(x)\Bigg(\frac{A(p^2) + A(l^2)}{2} \delta_{\mu, i}+\frac{C(p^2) + C(l^2)}{2} \delta_{\mu, 4} \Bigg),\\
\Gamma_{mod}(x) &= \frac{d_{1}}{\left(x+d_2\right)} + \frac{x}{\Lambda^2 + x} \left(\frac{\alpha(\mu)\beta_{0}}{4 \pi}\textrm{ln}\left(\frac{x}{\Lambda^2} + 1\right)\right)^{2 \delta}.
\end{align}
$\delta$ is the anomalous dimension of the ghost and the other parameters are the same as for the gluon dressing functions. All parameters of the models depending on the gauge group are listed in Ref.~\cite{Contant:2017gtz}. The gauge group dependent values of $d_1$ were fixed in \cite{Contant:2017gtz,Fischer:2014ata} and are given in \tref{tab:temperaturesAndChemPotentials}.

\section{Results}
\label{sec:results}

\begin{figure*}[tb]
 \includegraphics[width=0.4\textwidth]{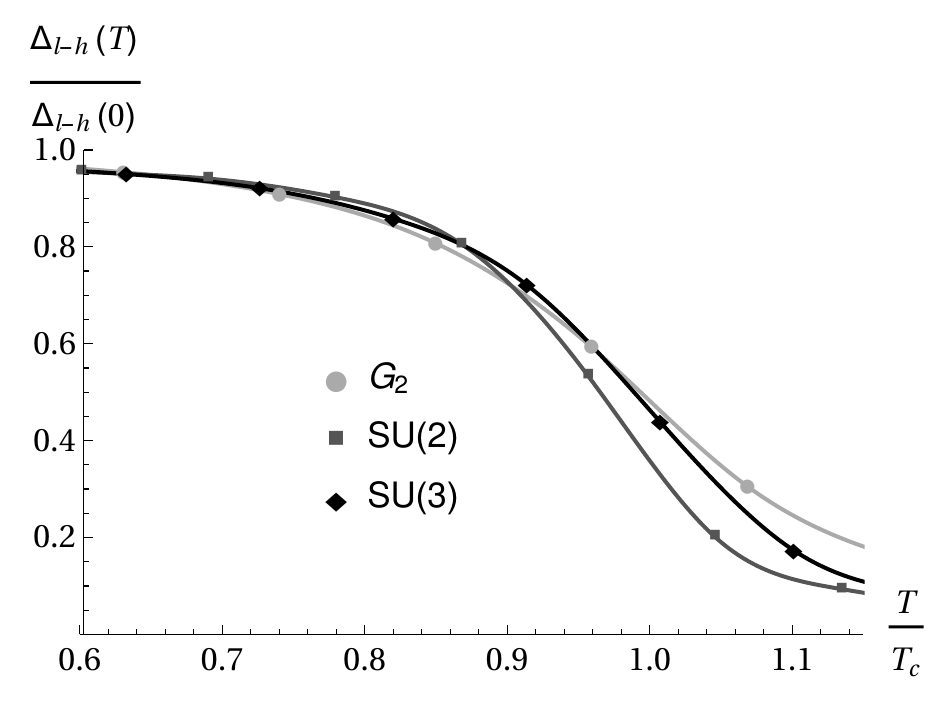}\hfill
 \includegraphics[width=0.4\textwidth]{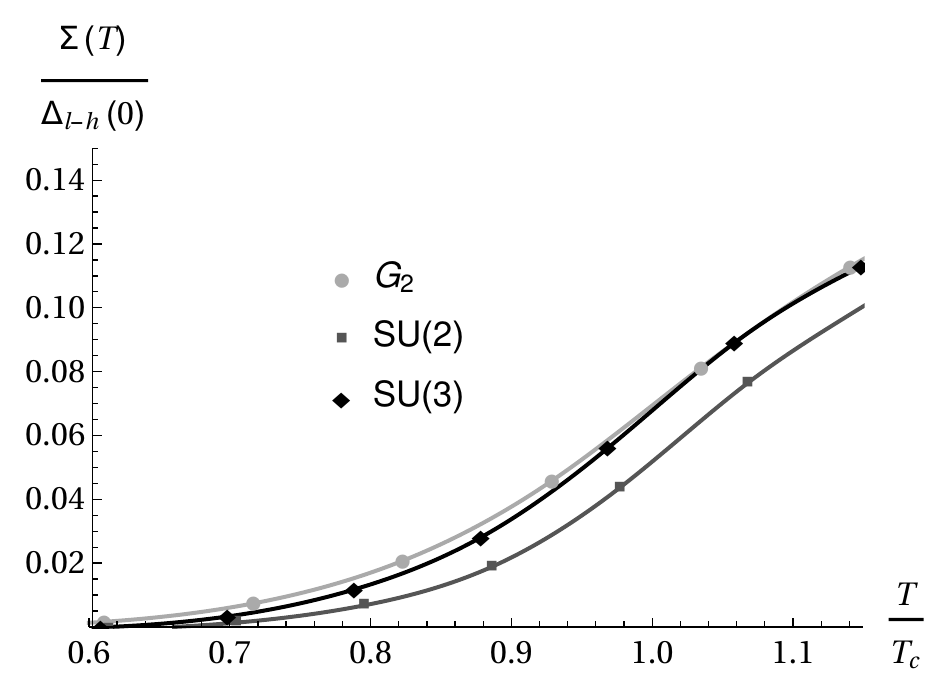}
 \caption{Chiral (\textit{left}) and dual (\textit{right}) condensates normalized to the vacuum chiral condensates for $N_f = 2$ with a bare quark mass of $m = 1.2 \,\text{MeV}$ at the renormalization point of $80\,\text{GeV}$.}
 \label{fig:Chidunf2fig}
\end{figure*}

We first recapitulate the results at $\mu=0$ from \cite{Contant:2017gtz}. In \tref{tab:temperaturesAndChemPotentials} the transition temperatures are listed and the condensates are shown in \fref{fig:Chidunf2fig}. The chiral and confinement/deconfinement transitions are very close to each other in all three cases. In general, we find a universal qualitative behavior of this truncation \cite{Contant:2017gtz}.

\begin{table}[b]
\begin{center}
\begin{tabular}{|l|l|l|l|}
 \hline
 & $SU(3)$ & $SU(2)$ & $G_2$   \\\hline \hline
 $d_1$ & $7.6\,\text{GeV}^2$ & $15\,\text{GeV}^2$ & $6.83\,\text{GeV}^2$ \\
 \hline
 $T_c(\mu=0)$ (chiral) & 194 MeV & 218 MeV & 153 MeV \\
 \hline
 $T_c(\mu=0)$ (deconfinement) & 201 MeV & 222 MeV & 157 MeV \\
 \hline
 $\mu_{CEP}$ (chiral) & 171 MeV & 200 MeV & 175 MeV \\
 \hline
 $T_{CEP}$ (chiral) & 158 MeV & 160 MeV & 115 MeV \\
 \hline
\end{tabular}
\caption{The crossover temperatures for $N_f=2$ at $\mu=0$ and the locations of the critical endpoints.}
\label{tab:temperaturesAndChemPotentials}
\end{center}
\end{table}

\label{sec:zeromu}

\begin{figure*}[tb]
 \includegraphics[width=0.5\textwidth]{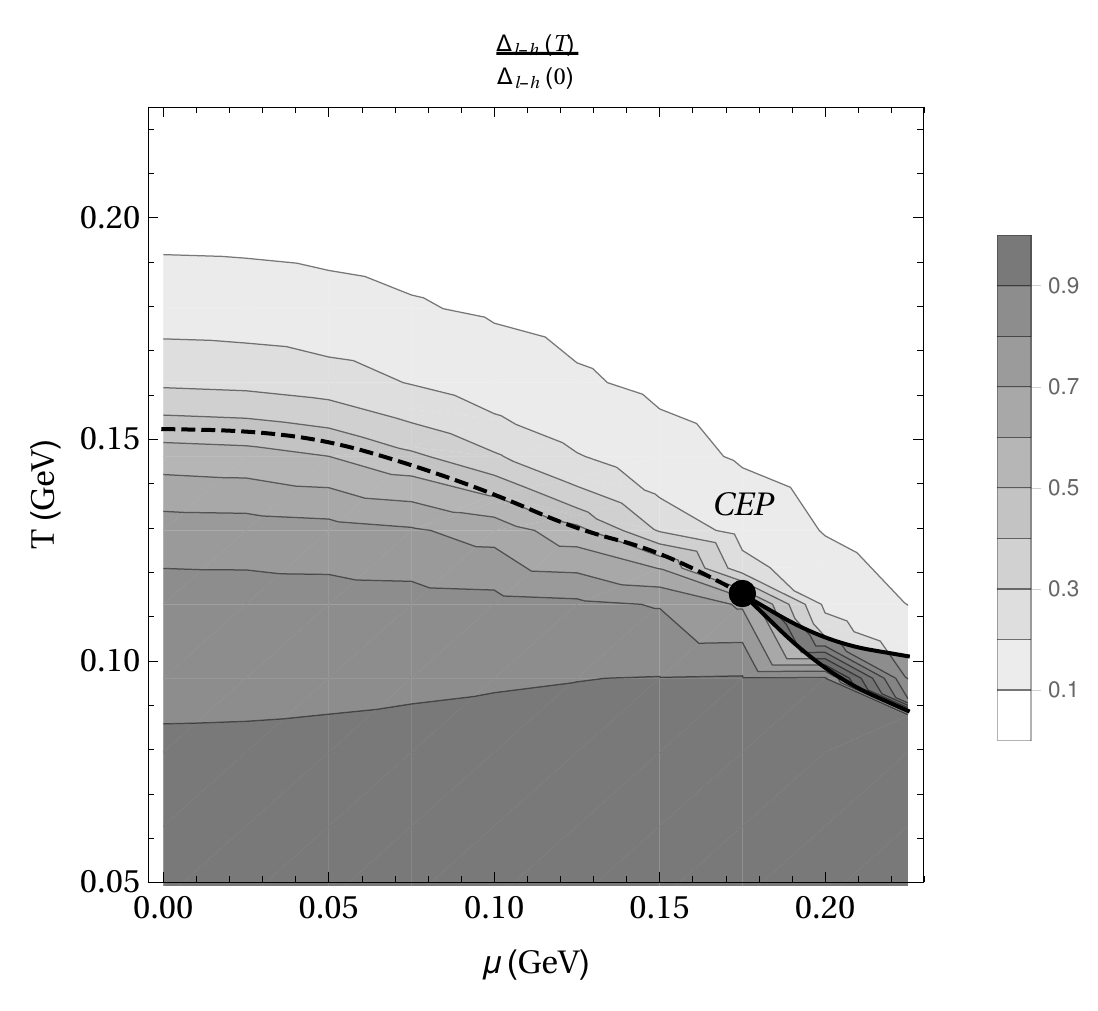}\hfill
 \includegraphics[width=0.5\textwidth]{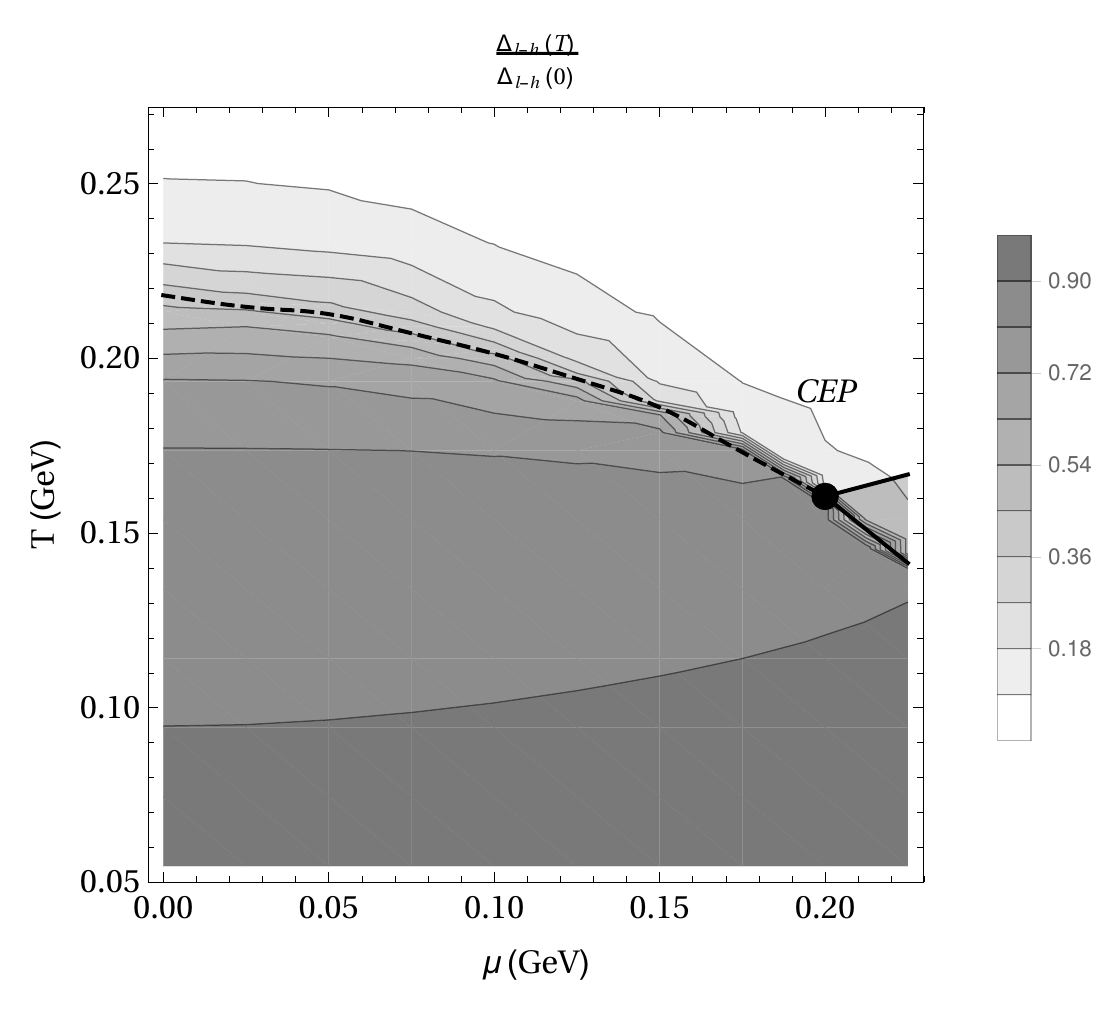}
 \caption{Chiral condensates of $G_2$ (\textit{left}) and $SU(2)$ (\textit{right}) for $N_f = 2$ normalized to the vacuum values. The dashed lines represent the crossover and the continuous lines the first order transition regions.}
 \label{fig:PhaseDiag}
\end{figure*}

Adding a light quark chemical potential we calculate the chiral crossover line until the critical endpoint beyond which it turns into a transition of first order, see \fref{fig:PhaseDiag}. Using two different initial conditions for the iterative solving procedure, we identify the spinodal lines of the first order regions.
This computation is done with a low resolution in $\mu$ and the accuracy of the position of the CEP within this truncation, see \tref{tab:temperaturesAndChemPotentials}, will be improved in the future.

\label{sec:phasediag}

\section{Summary}
\label{sec:summary}
We extended our analysis at $\mu=0$ \cite{Contant:2017gtz,Contant:2016ndj} of the universality of a DSE truncation scheme originally developed for $SU(3)$ \cite{Fischer:2012vc} by studying the chiral transition at $\mu>0$. Within the given truncation, we located the critical endpoints for the QCD-like theories with gauge groups $SU(2)$ and $G_2$. All three gauge groups show qualitatively the same behavior within this truncation. Before more detailed comparisons with lattice results are performed, we plan to include also diquarks in our calculations, which condense at $\mu = m_{\pi}/2$ according to chiral perturbation theory \cite{Kogut:2000ek} and lattice calculations \cite{Cotter:2012mb}. Also the resolution in $\mu$-direction will be improved. Further possible improvements include explicit calculations of the Yang-Mills sector or the quark-gluon vertex, which, however, are challenging projects on their own.

\section*{Acknowledgments}

Results have been obtained using the HPC clusters at the University of Graz.
Funding by the FWF (Austrian science fund) under Contract No. P 27380-N27 is gratefully acknowledged.

\bibliographystyle{utphys_mod}
\bibliography{literature_eQCD2017_CONTANT}

\providecommand{\href}[2]{#2}\begingroup\raggedright\begin{thebibliography}{10}

\bibitem{deForcrand:2010ys}
P.~de~Forcrand, {\em PoS} {\bfseries LAT2009} (2009) 010,
\href{http://arxiv.org/abs/1005.0539}{{\ttfamily arXiv:1005.0539 [hep-lat]}}.

\bibitem{Alkofer:2000wg}
R.~Alkofer and L.~von Smekal, {\em Phys. Rept.} {\bfseries 353} (2001) 281,
\href{http://arxiv.org/abs/hep-ph/0007355}{{\ttfamily arXiv:hep-ph/0007355}}.

\bibitem{Alkofer:2008nt}
R.~Alkofer, M.~Q. Huber, and K.~Schwenzer,
  \href{http://dx.doi.org/10.1016/j.cpc.2008.12.009}{{\em Comput. Phys.
  Commun.} {\bfseries 180} (2009) 965--976},
\href{http://arxiv.org/abs/0808.2939}{{\ttfamily arXiv:0808.2939 [hep-th]}}.

\bibitem{Berges:2000ew}
J.~Berges, N.~Tetradis, and C.~Wetterich, {\em Phys. Rept.} {\bfseries 363}
  (2002) 223--386,
\href{http://arxiv.org/abs/hep-ph/0005122}{{\ttfamily arXiv:hep-ph/0005122}}.

\bibitem{Gies:2006wv}
H.~Gies, \href{http://dx.doi.org/10.1007/978-3-642-27320-9_6}{{\em Lect.Notes
  Phys.} {\bfseries 852} (2012) 287--348},
\href{http://arxiv.org/abs/hep-ph/0611146}{{\ttfamily arXiv:hep-ph/0611146
  [hep-ph]}}.

\bibitem{Cyrol:2016tym}
A.~K. Cyrol, L.~Fister, M.~Mitter, J.~M. Pawlowski, and N.~Strodthoff,
  \href{http://dx.doi.org/10.1103/PhysRevD.94.054005}{{\em Phys. Rev.}
  {\bfseries D94} no.~5, (2016) 054005},
\href{http://arxiv.org/abs/1605.01856}{{\ttfamily arXiv:1605.01856 [hep-ph]}}.

\bibitem{Cyrol:2017ewj}
A.~K. Cyrol, M.~Mitter, J.~M. Pawlowski, and N.~Strodthoff,
\href{http://arxiv.org/abs/1706.06326}{{\ttfamily arXiv:1706.06326 [hep-ph]}}.

\bibitem{Kogut:2000ek}
J.~B. Kogut, M.~A. Stephanov, D.~Toublan, {\em et~al.},
  \href{http://dx.doi.org/10.1016/S0550-3213(00)00242-X}{{\em Nucl. Phys.}
  {\bfseries B582} (2000) 477--513},
\href{http://arxiv.org/abs/hep-ph/0001171}{{\ttfamily arXiv:hep-ph/0001171
  [hep-ph]}}.

\bibitem{Holland:2003jy}
K.~Holland, P.~Minkowski, M.~Pepe, and U.~J. Wiese,
  \href{http://dx.doi.org/10.1016/S0550-3213(03)00571-6}{{\em Nucl. Phys.}
  {\bfseries B668} (2003) 207--236},
\href{http://arxiv.org/abs/hep-lat/0302023}{{\ttfamily arXiv:hep-lat/0302023
  [hep-lat]}}.

\bibitem{Maas:2012wr}
A.~Maas, L.~von Smekal, B.~Wellegehausen, and A.~Wipf,
  \href{http://dx.doi.org/10.1103/PhysRevD.86.111901}{{\em Phys. Rev.}
  {\bfseries D86} (2012) 111901},
\href{http://arxiv.org/abs/1203.5653}{{\ttfamily arXiv:1203.5653 [hep-lat]}}.

\bibitem{Cotter:2012mb}
S.~Cotter, P.~Giudice, S.~Hands, and J.-I. Skullerud,
  \href{http://dx.doi.org/10.1103/PhysRevD.87.034507}{{\em Phys.Rev.}
  {\bfseries D87} (2013) 034507},
\href{http://arxiv.org/abs/1210.4496}{{\ttfamily arXiv:1210.4496 [hep-lat]}}.

\bibitem{Boz:2013rca}
T.~Boz, S.~Cotter, L.~Fister, D.~Mehta, and J.-I. Skullerud,
  \href{http://dx.doi.org/10.1140/epja/i2013-13087-6}{{\em Eur. Phys. J.}
  {\bfseries A49} (2013) 87},
\href{http://arxiv.org/abs/1303.3223}{{\ttfamily arXiv:1303.3223 [hep-lat]}}.

\bibitem{Braguta:2016cpw}
V.~V. Braguta, E.~M. Ilgenfritz, A.~{\relax Yu}. Kotov, A.~V. Molochkov, and
  A.~A. Nikolaev, \href{http://dx.doi.org/10.1103/PhysRevD.94.114510}{{\em
  Phys. Rev.} {\bfseries D94} no.~11, (2016) 114510},
\href{http://arxiv.org/abs/1605.04090}{{\ttfamily arXiv:1605.04090 [hep-lat]}}.

\bibitem{Maas:2007af}
A.~Maas and S.~Olejnik,
  \href{http://dx.doi.org/10.1088/1126-6708/2008/02/070}{{\em JHEP} {\bfseries
  02} (2008) 070},
\href{http://arxiv.org/abs/0711.1451}{{\ttfamily arXiv:0711.1451 [hep-lat]}}.

\bibitem{Sternbeck:2007ug}
A.~Sternbeck, L.~von Smekal, D.~Leinweber, and A.~Williams, {\em PoS}
  {\bfseries LAT2007} (2007) 340,
\href{http://arxiv.org/abs/0710.1982}{{\ttfamily arXiv:0710.1982 [hep-lat]}}.

\bibitem{Cucchieri:2007zm}
A.~Cucchieri, T.~Mendes, O.~Oliveira, and P.~J. Silva,
  \href{http://dx.doi.org/10.1103/PhysRevD.76.114507}{{\em Phys. Rev.}
  {\bfseries D76} (2007) 114507},
\href{http://arxiv.org/abs/0705.3367}{{\ttfamily arXiv:0705.3367 [hep-lat]}}.

\bibitem{Fischer:2010fx}
C.~S. Fischer, A.~Maas, and J.~A. Mueller,
  \href{http://dx.doi.org/10.1140/epjc/s10052-010-1343-1}{{\em Eur. Phys. J.}
  {\bfseries C68} (2010) 165--181},
\href{http://arxiv.org/abs/1003.1960}{{\ttfamily arXiv:1003.1960 [hep-ph]}}.

\bibitem{Maas:2011ez}
A.~Maas, J.~M. Pawlowski, L.~von Smekal, and D.~Spielmann,
  \href{http://dx.doi.org/10.1103/PhysRevD.85.034037}{{\em Phys.Rev.}
  {\bfseries D85} (2012) 034037},
\href{http://arxiv.org/abs/1110.6340}{{\ttfamily arXiv:1110.6340 [hep-lat]}}.

\bibitem{Contant:2017gtz}
R.~Contant and M.~Q. Huber,
\href{http://arxiv.org/abs/1706.00943}{{\ttfamily arXiv:1706.00943 [hep-ph]}}.

\bibitem{Bilgici:2008qy}
E.~Bilgici, F.~Bruckmann, C.~Gattringer, and C.~Hagen,
  \href{http://dx.doi.org/10.1103/PhysRevD.77.094007}{{\em Phys. Rev.}
  {\bfseries D77} (2008) 094007},
\href{http://arxiv.org/abs/0801.4051}{{\ttfamily arXiv:0801.4051 [hep-lat]}}.

\bibitem{Synatschke:2007bz}
F.~Synatschke, A.~Wipf, and C.~Wozar,
  \href{http://dx.doi.org/10.1103/PhysRevD.75.114003}{{\em Phys. Rev.}
  {\bfseries D75} (2007) 114003},
\href{http://arxiv.org/abs/hep-lat/0703018}{{\ttfamily arXiv:hep-lat/0703018}}.

\bibitem{Fischer:2009wc}
C.~S. Fischer, \href{http://dx.doi.org/10.1103/PhysRevLett.103.052003}{{\em
  Phys. Rev. Lett.} {\bfseries 103} (2009) 052003},
\href{http://arxiv.org/abs/0904.2700}{{\ttfamily arXiv:0904.2700 [hep-ph]}}.

\bibitem{Fischer:2012vc}
C.~S. Fischer and J.~Luecker,
  \href{http://dx.doi.org/10.1016/j.physletb.2012.11.054}{{\em Phys.Lett.}
  {\bfseries B718} (2013) 1036--1043},
\href{http://arxiv.org/abs/1206.5191}{{\ttfamily arXiv:1206.5191 [hep-ph]}}.

\bibitem{Huber:2011xc}
M.~Q. Huber and M.~Mitter,
  \href{http://dx.doi.org/10.1016/j.cpc.2012.05.019}{{\em Comput.Phys.Commun.}
  {\bfseries 183} (2012) 2441--2457},
\href{http://arxiv.org/abs/1112.5622}{{\ttfamily arXiv:1112.5622 [hep-th]}}.

\bibitem{Fischer:2009gk}
C.~S. Fischer and J.~A. M\"uller,
  \href{http://dx.doi.org/10.1103/PhysRevD.80.074029}{{\em Phys. Rev.}
  {\bfseries D80} (2009) 074029},
\href{http://arxiv.org/abs/0908.0007}{{\ttfamily arXiv:0908.0007 [hep-ph]}}.

\bibitem{Fischer:2014ata}
C.~S. Fischer, J.~Luecker, and C.~A. Welzbacher,
  \href{http://dx.doi.org/10.1103/PhysRevD.90.034022}{{\em Phys. Rev.}
  {\bfseries D90} no.~3, (2014) 034022},
\href{http://arxiv.org/abs/1405.4762}{{\ttfamily arXiv:1405.4762 [hep-ph]}}.

\bibitem{Contant:2016ndj}
R.~Contant and M.~Q. Huber,
  \href{http://dx.doi.org/10.1051/epjconf/201713713003}{{\em EPJ Web Conf.}
  {\bfseries 137} (2017) 13003},
\href{http://arxiv.org/abs/1612.00691}{{\ttfamily arXiv:1612.00691 [hep-ph]}}.

\end{thebibliography}\endgroup

\end{document}